\documentclass[jfm,aps,preprint,showpacs,preprintnumbers,amsmath,amssymb,secnumroman]{revtex4}

\usepackage{graphicx}% Include figure files
\usepackage{dcolumn}% Align table columns on decimal point
\usepackage{bm}% bold math

\newcommand{\beq}{\begin{equation}}
\newcommand{\eeq}{\end{equation}}
\newcommand{\beqa}{\begin{eqnarray}}
\newcommand{\eeqa}{\end{eqnarray}}
\newcommand{\vc}[1]{\mbox{\boldmath $#1$}}

\newcommand{\vol}[1]{{\bf #1}}

\begin{document}

%\preprint{APS/123-QED}

\title{Swimming of a circular disk at low Reynolds number}% Force line breaks with \\

\author{B. U. Felderhof}

 \email{ufelder@physik.rwth-aachen.de}
\affiliation{Institut f\"ur Theoretische Physik A\\ RWTH Aachen\\
Templergraben 55\\52056 Aachen\\ Germany\\
}%

\date{\today}% It is always \today, today,

\begin{abstract}
The swimming of a circular disk at low Reynolds number is studied for distortion waves along its two planar surfaces with wavelength much smaller than the size of the disk. The calculation is based on an extension  of Taylor's work for a planar sheet. It is shown that in general the disk performs both translational and rotational swimming, resulting in a circular orbit.
\end{abstract}

\pacs{47.15.G-, 47.63.mf, 47.63.Gd, 87.17.Jj}% PACS, the Physics and Astronomy
\maketitle
\section{\label{I}Introduction}

Taylor's seminal paper on the swimming of microscopic organisms (Taylor 1951) generated a wide range of studies of swimming at low Reynolds number (Shapere $\&$ Wilczek 1989a,b; Felderhof $\&$ Jones 1994a,b; Lauga $\&$ Powers 2009; Garstecki $\&$ Cieplak 2009; Swan et al. 2011). Taylor considered the swimming of a planar sheet with sinusoidal transverse undulation. Soon after, Lighthill (1952) published a similar calculation for a squirming sphere. The theory is based on Stokes' equations for flow velocity and pressure of a viscous incompressible fluid (Happel $\&$ Brenner 1973). A no-slip boundary condition applies at the interface of body and fluid. The swimming speed and the rate of dissipation are calculated in perturbation theory to second order in the surface displacements.
Taylor's calculation is easily extended to the swimming of a planar slab. The independent distortions of the upper and lower surface of the slab allows a wider class of motions.  Taylor's model is recovered in the limit of a thin slab with parallel motion of the two surfaces.

Taylor's calculation was extended by Reynolds (1965), Tuck (1968), Blake (1971), and Childress (1981) to include squirming motion of the planar sheet. We show in the following that their calculation of the swimming velocity was incorrect. For the more general motion the second order flow velocity differs on the two sides of the sheet, and this was not taken into account appropriately. In Sec. 3 of this article we consider the sheet with a superposition of undulating and squirming distortions in detail. For pure undulation and pure squirming there is no problem, and the sheet has a well-defined swimming velocity. However, for the superposition of the two modes of flow the second order velocity shows a discontinuity at the undisplaced sheet due to interference of flows with different symmetry on the two sides. The discontinuity was discussed by Reynolds (1965), but he interpreted the situation as a pumping mechanism.

We resolve the difficulty by considering a finite system, in particular a circular disk. It then becomes evident that the difference in surface velocity on either side corresponds to a torque exerted on the fluid. The torque must be canceled by adding a Stokes flow with no-slip boundary condition on the disk but with rotation of the fluid at infinity (Felderhof $\&$ Jones 1994a). Hence one infers a net rotational swimming velocity in the laboratory frame.

Subsequently we study the mean translational and rotational swimming of a circular disk in more detail. Since both sides of the disk can have independent surface distortions, a wide range of flow situations is possible. For the optimal stroke the efficiency, defined from the ratio of speed and power, is a factor $\sqrt{2}$ larger than for Taylor's undulating sheet.

For the optimal stroke the mean rotational swimming velocity vanishes. For the general stroke the combined translational and rotational swimming leads to a circular orbit. At any point the disk is perpendicular to the plane of the circle and tangential to the circle. Consequently, the possibility of slightly different strokes on the two sides of its body allows a swimming flat microorganism (Opalina or Paramecium) to change direction.

\section{\label{2}Taylor's swimming sheet}

We consider a planar slab immersed in a viscous
incompressible fluid of shear viscosity $\eta$. The slab at rest is bounded by two planes at distance $2d$. We choose Cartesian coordinates $x,y,z$ such that the upper plane is at $y=d$ and the lower at $y=-d$. The rest shape of the slab is denoted as $S_0$. We shall consider a prescribed time-dependent shape $S(t)$ leading to swimming motion in the $x$ direction. The distortions are decomposed into $S_+(t)$ for the upper surface and $S_-(t)$ for the lower surface. The fluid is set in motion by the time-dependent distortions of the
slab. At low Reynolds
number and on a slow time scale the flow velocity
$\vc{v}(\vc{r},t)$ and the pressure $p(\vc{r},t)$ in the rest frame satisfy the
Stokes equations
\begin{equation}
\label{2.1}\eta\nabla^2\vc{v}-\nabla p=0,\qquad\nabla\cdot\vc{v}=0.
\end{equation}

The surface displacement
$\vc{\xi}(\vc{s},t)$ is defined as the vector distance
\begin{equation}
\label{2.2}\vc{\xi}=\vc{s}'-\vc{s}
\end{equation}
of a point $\vc{s}'$ on the displaced surface $S(t)$ from the
point $\vc{s}$ on the slab $S_0$. We decompose $\vc{s}$ into $\vc{s}_+=(x,d,z)$ for the upper surface and $\vc{s}_-=(x,-d,z)$ for the lower surface. The fluid
velocity $\vc{v}(\vc{r},t)$ is required to satisfy
\begin{equation}
\label{2.3}\vc{v}(\vc{s}+\vc{\xi}(\vc{s},t))=\frac{\partial\vc{\xi}(\vc{s},t)}{\partial t}.
\end{equation}
This amounts to a no-slip boundary condition. The displacement is decomposed into $\vc{\xi}_\pm(t)$ for the upper and lower surface. We consider displacements which do not depend on $z$ and take the form $\vc{\xi}_\pm=(\xi_{\pm x}(x,t),\xi_{\pm y}(x,t),0)$. As a consequence the flow velocity $\vc{v}$ and pressure $p$ do not depend on $z$, and the problem is effectively two-dimensional.

First we recall Taylor's well-known calculation of the swimming of a thin planar sheet due to a running wave of surface undulations (Taylor 1951). The sheet  can be regarded as the limiting case of a slab with $d\rightarrow 0$ and $S_+(t)$ and $S_-(t)$ moving in parallel so that $\vc{\xi}_+(t)=\vc{\xi}_-(t)$.
Taylor considered transverse oscillations of the sheet with surface distortions described by the running wave
\begin{equation}
\label{2.4}\vc{\xi}(x,t)=A\sin(kx-\omega t)\vc{e}_y,
\end{equation}
with amplitude $A$, positive wavenumber $k$, and positive frequency $\omega$, so that the wave propagates in the positive $x$ direction. He derived the flow pattern from a stream function $\psi(x,y,t)$, defined such that the components of velocity are
 \begin{equation}
\label{2.5}v_x=u=\frac{\partial\psi}{\partial y},\qquad v_y=v=-\frac{\partial\psi}{\partial x}.
\end{equation}
(We use a different sign convention.) Stokes' equations Eq. (1) are satisfied provided that the stream function satisfies the biharmonic equation
 \begin{equation}
\label{2.6}\nabla^4\psi=0.
\end{equation}
To terms linear in $A$ the stream function is found to be given by
 \begin{equation}
\label{2.7}\psi_1=A\frac{\omega}{k}(1+k|y|)e^{-k|y|}\sin(kx-\omega t).
\end{equation}
The first order velocity components are
\begin{equation}
\label{2.8}u_1=-A\omega kye^{-k|y|}\sin(kx-\omega t),\qquad v_1=-A\omega (1+k|y|)e^{-k|y|}\cos(kx-\omega t).
\end{equation}
The corresponding pressure disturbance is given by
 \begin{equation}
\label{2.9}p_1=\mp2A\eta\omega ke^{-k|y|}\cos(kx-\omega t),
\end{equation}
where the upper (lower) sign corresponds to the upper (lower) half-space. The velocity components satisfy the symmetry relations
 \begin{equation}
\label{2.10}u_1(x,y,t)=-u_1(x,-y,t),\qquad
v_1(x,y,t)=v_1(x,-y,t).
\end{equation}
We call this mirror-inverse symmetry. It corresponds to reflection of the vector in the plane $y=0$ and inversion.

Taylor showed that to second order in the amplitude $A$ the sheet swims in the negative $x$ direction with speed
\begin{equation}
\label{2.11}|U_2|=\frac{1}{2}\omega kA^2.
\end{equation}
Alternatively this result can be derived from the observation that the second order time-averaged flow $\overline{\vc{v}}_2,\overline{p}_2$ must satisfy the Stokes equations Eq. (1) with the boundary condition (Felderhof $\&$ Jones 1994a)
\begin{equation}
\label{2.12}\overline{\vc{v}}_2(\vc{s})=-\overline{(\vc{\xi}\cdot\nabla)\vc{v}_1}\big|_{y=0},
\end{equation}
where the overline indicates averaging over a period $T=2\pi/\omega$. The expression on the right-hand side follows from the boundary condition Eq. (3). It takes the value $|U_2|$, so that Eq. (1) is solved by
\begin{equation}
\label{2.13}\overline{\vc{v}}_2(\vc{r})=|U_2|\vc{e}_x,\qquad\overline{p}_2=0.
\end{equation}
This implies that the swimming velocity of the sheet in the laboratory frame is $\vc{U}_2=-|U_2|\vc{e}_x$.

The required power equals the rate of dissipation of energy in the fluid. Taylor calculated this from the work done per unit area against the stress $\vc{\sigma}=\eta(\nabla\vc{v}+\widetilde{\nabla\vc{v}})-p\vc{I}$. The mean rate of dissipation per unit area is to second order (Felderhof $\&$ Jones 1994a)
\begin{equation}
\label{2.14}\overline{D}_2=-2\overline{\vc{v}_1\cdot\vc{\sigma}_1\cdot\vc{e}_y}\big|_{y=0}.
\end{equation}
The factor 2 arises from the fact that both sides of the sheet must be considered. It turns out that the viscous stress does not contribute, so that Eq. (14) reduces to
\begin{equation}
\label{2.15}\overline{D}_2=2\overline{\frac{\partial\xi_{y}}{\partial t}p_1}\big|_{y=0}.
\end{equation}
Substituting from Eqs. (4) and (9) one finds Taylor's result
\begin{equation}
\label{2.16}\overline{D}_2=2\eta\omega^2 kA^2.
\end{equation}

We define the dimensionless efficiency as
\begin{equation}
\label{2.17}E_2=4\eta\omega\frac{|U_2|}{\overline{D}_2}.
\end{equation}
In the present case this equals unity. We have shown elsewhere (Felderhof 2009) how the efficiency is affected by walls at $y=\pm L$.

Taylor's calculation can be generalized easily to a slab of thickness $2d$ with surface displacements $\vc{\xi}_{\pm}(x,t)$ of the upper and lower surface at $y=\pm d$, both given by Eq. (4). The flow pattern in the upper half-space is simply shifted by $d$ in the positive $y$ direction, and in the lower half-space by $d$ in the negative $y$ direction. In the above expressions $y$ must be replaced by $y\mp d$. The speed and the power are the same as for the infinitesimally thin sheet.

\section{\label{III}Undulating and squirming sheet}

Taylor's calculation for a thin planar sheet has been generalized to include squirming distortions, besides the transverse motions. Tuck (1968) called such motions longitudinal and calculated the corresponding swimming velocity, including corrections due to inertia. The motion is also included in the derivations of Reynolds (1965), Blake (1971), and Childress (1981). We begin the discussion by considering the squirming motion by itself.

We consider longitudinal oscillations of the sheet with surface distortions described by the running wave
\begin{equation}
\label{3.1}\vc{\xi}(x,t)=B\sin(kx-\omega t+\alpha)\vc{e}_x,
\end{equation}
with a phase shift $\alpha$ with respect to Eq. (4). We assume that $B$ is positive and consider values of $\alpha$ in the range $-\pi\leq\alpha\leq\pi$. The corresponding stream function is given by
 \begin{equation}
\label{3.2}\psi_1=-B\omega ye^{-k|y|}\cos(kx-\omega t+\alpha).
\end{equation}
The first order velocity components are
\begin{equation}
\label{3.3}u_1=B\omega(-1+k|y|e^{-k|y|})\cos(kx-\omega t+\alpha),\qquad v_1=-B\omega kye^{-k|y|}\sin(kx-\omega t+\alpha).
\end{equation}
The corresponding pressure disturbance is given by
 \begin{equation}
\label{3.4}p_1=-2B\eta\omega ke^{-k|y|}\sin(kx-\omega t+\alpha).
\end{equation}
The velocity components satisfy the symmetry relations
 \begin{equation}
\label{3.5}u_1(x,y,t)=u_1(x,-y,t),\qquad
v_1(x,y,t)=-v_1(x,-y,t).
\end{equation}
We call this mirror symmetry.

For the linear combination of the displacements in Eqs. (4) and (18) we find for the velocity to be used in the second order boundary condition as in Eq. (12)
\begin{eqnarray}
\label{3.6}\vc{v}_2(\vc{s}_\pm,t)&=&-(\vc{\xi}\cdot\nabla)\vc{v}_1\big|_{y=0\pm}\nonumber\\
&=&\omega k\big[A^2 \sin^2\varphi-B^2\sin^2(\varphi+\alpha)\mp2AB\sin\varphi\cos(\varphi+\alpha)\big]\vc{e}_x,
\end{eqnarray}
where we have abbreviated
 \begin{equation}
\label{3.7}\varphi=kx-\omega t.
\end{equation}
Thus in general the second order fluid velocity tends to different values on either side of the sheet.
Averaging over a period we find
 \begin{equation}
\label{3.8}\overline{\vc{v}}_2(\vc{s}_\pm)=\frac{1}{2}\omega k\big[A^2-B^2\pm 2AB\sin\alpha\big]\vc{e}_x,
\end{equation}
with different values on both sides of the sheet if $\alpha>0$ and both $A$ and $B$ differ from zero. Blake (1971) and Childress (1981) explicitly consider only the upper side of the sheet, and arrive at the result for the swimming velocity of the sheet $\overline{\vc{U}}_{2BC}=\overline{U}_{2BC}\vc{e}_x$ with
 \begin{equation}
\label{3.9}\overline{U}_{2BC}=\frac{1}{2}\omega k\big[-A^2+B^2-2AB\sin\alpha\big],
\end{equation}
(Blake considered the case $\alpha=\pi/2$ in his Eq. (40)). This would allow one to use the phase angle $\alpha$ to optimize the speed for fixed $A$ and $B$. However, the result is wrong, since in the derivation only the upper sign in Eq. (25) was considered.

The rate of dissipation turns out to be the same on both sides of the sheet. Calculating as in Eq. (14) one finds for the mean rate of dissipation per unit area
\begin{equation}
\label{3.10}\overline{D}_2=2\eta\omega^2 k(A^2+B^2),
\end{equation}
independent of the phase angle $\alpha$.

The solution of the second order flow problem for the infinite sheet is not obvious. In order to understand the physical situation we must regard the sheet as an idealization of the problem for a finite system.

\section{\label{IV}Circular disk}

Specifically we consider a circular disk of radius $a$ in the horizontal $x,z$ plane and of height $2d$ in the $y$ direction. The origin is taken at the center of the disk. We assume that the disk is thin with $d$ much smaller than $a$. In the theory of elasticity this would be called a plate (Green $\&$ Zerna 1968), and it would be surrounded by vacuum, rather than by a viscous incompressible fluid. We assume that the upper and lower surface perpendicular to the $y$ axis are distorted in phase with equal displacements on both sides of the disk,
\begin{equation}
\label{4.1}\vc{\xi}(\vc{s}_+,t)=\vc{\xi}(\vc{s}_-,t)=A\sin(kx-\omega t)\vc{e}_y+B\sin(kx-\omega t+\alpha)\vc{e}_x.
\end{equation}
The fluid moves on account of the no-slip boundary condition.
We consider distortions such that the wavelength $\lambda=2\pi/k$ is much larger than the thickness $2d$ and much smaller than the horizontal dimensions. The boundary conditions
  \begin{equation}
\label{4.2}\vc{v}_2(\vc{r},t)\big|_{y=\pm d+}=\vc{v}_2(\vc{s}_\pm,t),
\end{equation}
with right-hand side given by Eq. (23), must be applied within the upper and lower circular areas $S_\pm$ parallel to the $xz$ plane.
Averaging over a period $T=2\pi/\omega$ this becomes
  \begin{equation}
\label{4.3}\overline{\vc{v}}_2(\vc{s}_\pm)=g_\pm\vc{e}_x,\qquad \vc{s}_\pm\in S_\pm,
\end{equation}
with constants $g_\pm$ given by the factor in Eq. (25). The symmetric part is identified as minus the translational swimming velocity
  \begin{equation}
\label{4.4}\overline{U}_2=-\frac{1}{2}(g_++g_-)=\frac{1}{2}\omega k(-A^2+B^2).
\end{equation}
The antisymmetric part $\frac{1}{2}(g_+-g_-)$ gives rise to a rotational swimming velocity. The difference in fluid velocity on both sides of the disk corresponds to a uniform force dipole layer. At large distances the flow in the external fluid tends to the flow generated by a point torque at the origin. From the force dipole layer the torque acting on the fluid is calculated as
  \begin{equation}
\label{4.5}\vc{N}=-\pi\eta(g_+-g_-)a^2\vc{e}_z=-2\pi\eta\omega ka^2AB\sin\alpha\;\vc{e}_z.
\end{equation}
Note that for a uniform force dipole layer only the area covered is relevant, so that a similar calculation would work for an elliptical disk, for example.
At large distances the flow tends to
  \begin{equation}
\label{4.6}\hat{\vc{v}}_2(\vc{r})\approx\frac{1}{8\pi\eta}\vc{N}\times\frac{\vc{r}}{r^3},
\qquad\hat{p}_2(\vc{r})\approx 0,\qquad\mathrm{as}\;r\rightarrow\infty.
\end{equation}
Outside the disk the flow velocity $\hat{\vc{v}}_2(\vc{r})$ satisfies Laplace's equation and the pressure $\hat{p}_2$ vanishes. The flow velocity is identical with the vector potential
of a magnet in the shape of the disk with magnetization in the plane of the disk.
In order to find the complete second order flow pattern at large distances we must add the Stokes flow which vanishes at the surface of the disk (Felderhof $\&$ Jones 1994a), and which cancels the term shown in Eq. (33). This flow has the asymptotic behavior
  \begin{equation}
\label{4.7}\vc{v}^{\mathrm{St}}_2(\vc{r})\approx -\overline{\vc{\Omega}}_2\times\vc{r}-\frac{1}{8\pi\eta}\vc{N}\times\frac{\vc{r}}{r^3},
\qquad p^{\mathrm{St}}_2(\vc{r})\approx 0,\qquad\mathrm{as}\;r\rightarrow\infty,
\end{equation}
where $\overline{\vc{\Omega}}_2$ is related to $\vc{N}$ by (Happel $\&$ Brenner 1973)
  \begin{equation}
\label{4.8}\vc{N}=-\frac{32}{3}\eta a^3\overline{\vc{\Omega}}_2.
\end{equation}
The complete mean second order flow tends to
 \begin{equation}
\label{4.9}\overline{\vc{v}}_2(\vc{r})\approx-\overline{\vc{\Omega}}_2\times\vc{r}-\overline{U}_2\vc{e}_x,
\qquad\overline{p}_2(\vc{r})\approx 0,\qquad\mathrm{as}\;r\rightarrow\infty.
\end{equation}
This identifies $\overline{\vc{\Omega}}_2$ as the mean rotational swimming velocity of the disk in the laboratory frame. From Eqs. (32) and (35) we find
  \begin{equation}
\label{4.10}\overline{\vc{\Omega}}_2=\frac{3\pi}{16a}\;\omega k AB\sin\alpha\;\vc{e}_z.
\end{equation}
This shows that on time average the disk rotates steadily about a diameter perpendicular to the direction of translational swimming. Since the translational swimming velocity is always parallel to the plane of the disk this implies that on  a slow time scale the center of the disk runs through a circular orbit in the laboratory frame. The circle has radius $|U_2|/|\Omega_2|$ and is traversed in time $2\pi/|\Omega_2|$. At any point the disk is orthogonal and tangential to the circle.

\section{\label{V}Twofold strokes}

In this section we consider more complicated modes of motion in which the disk has different distortions on the two sides. We can idealize the disk as an infinite slab with surface displacements of the upper and lower surface given by
\begin{eqnarray}
\label{5.1}\vc{\xi}(\vc{s}_+,t)&=&A\sin(kx-\omega t)\vc{e}_y+B\sin(kx-\omega t+\alpha)\vc{e}_x,\nonumber\\
\vc{\xi}(\vc{s}_-,t)&=&C\sin(kx-\omega t+\beta)\vc{e}_y+D\sin(kx-\omega t+\beta+\gamma)\vc{e}_x.
\end{eqnarray}
The no-slip boundary condition Eq. (3) is imposed on both sides. The flow in the upper half-space is the same as before, and in the lower half-space it is modified only by amplitudes and phase shifts. Hence the mean second order flow velocity at the two surfaces becomes
 \begin{eqnarray}
\label{5.2}\overline{\vc{v}}_2(\vc{s}_+)&=&\frac{1}{2}\omega k\big[A^2-B^2+2AB\sin\alpha\big]\vc{e}_x,\nonumber\\
\overline{\vc{v}}_2(\vc{s}_-)&=&\frac{1}{2}\omega k\big[C^2-D^2-2CD\sin\gamma\big]\vc{e}_x.
\end{eqnarray}
Therefore the mean translational swimming velocity of the disk is
 \begin{equation}
\label{5.3}\overline{\vc{U}}_2=\frac{1}{4}\omega k\big[-A^2+B^2-C^2+D^2-2AB\sin\alpha+2CD\sin{\gamma}\big]\;\vc{e}_x,
\end{equation}
and the mean rotational swimming velocity is
  \begin{equation}
\label{5.4}\overline{\vc{\Omega}}_2=\frac{3\pi}{64a}\;\omega k\big[A^2-B^2-C^2+D^2+2AB\sin\alpha+2CD\sin{\gamma}\big]\;\vc{e}_z.
\end{equation}
The mean rate of dissipation per unit area is given by
\begin{equation}
\label{5.5}\overline{D}_2=\eta\omega^2 k(A^2+B^2+C^2+D^2),
\end{equation}
in generalization of Eq. (27).

A case of particular interest corresponds to $C=-A,\;D=-B,\;\beta=0,\;\gamma=\alpha$. This is a squeezing mode in which the lower surface is distorted in the opposite direction to the upper one. Clearly the two swimming velocities are the same as for the stretching mode studied earlier. For the symmetric squeezing mode corresponding to $C=-A,\;D=B,\;\beta=0,\;\gamma=\alpha$ the translational swimming velocity is given by Eq. (26), corresponding to Blake's Eq. (36), and the rotational swimming velocity vanishes. Blake did not distinguish the slab with symmetric squeezing mode, shown in his Fig. 1, from the two-dimensional waving infinitesimally thin sheet considered by Taylor and Tuck.

The tip of the displacement vector at any point of the surface, as given by Eq. (38), in general runs through an elliptical orbit as a function of time. The semi-axes, the ellipticity, and the tilt angle can be calculated as Stokes parameters (Bohren $\&$ Huffman 1983). The parameters in Eq. (38) can be varied to optimize the swimming speed for given power. This corresponds to maximization of the efficiency $E_2$, defined in Eq. (17). In particular for $C=A,\;D=B,\gamma=-\alpha$, or for $C=-A,\;D=-B,\gamma=-\alpha$,  the mean swimming velocity becomes equal to the expression of Blake and Childress given by Eq. (26). For such strokes the efficiency $E_2$ is maximal for $A=\pm B,\;\alpha=\pi/2$, and equal to $\sqrt{2}$, substantially larger than unity, the value for the Taylor mode. For the optimum stroke the rotational swimming velocity vanishes.

\section{\label{VI}Discussion}

In the above we have studied the swimming of a circular disk caused by synchronized running distortion waves on the upper and lower surface of the disk. For asymmetric situations the disk performs both translational and rotational swimming.

The swimming of a slab generalizes the swimming of a sheet studied originally by Taylor (1951). His calculation has been extended to swimming with account of fluid inertia (Reynolds 1965, Tuck 1968), and to swimming in confined geometry (Katz 1974). Such studies could also be performed for a slab. The results can be applied to the swimming of a circular disk.

The calculations can be extended to the case of an elliptical disk with distortion waves on either side running in the direction of the long or short semi-axis. The mean translational swimming velocity remains the same as Eq. (40). The coefficient of the mean rotational velocity in Eq. (41) is modified and can be calculated from the known expressions for the rotational friction coefficient of an ellipsoid (Kim $\&$ Karrila 1991).

In the same article Taylor studied also the swimming of two sheets in parallel. Elsewhere we have studied the efficiency of swimming in this situation (Felderhof 2012). Similarly one could consider the swimming of two parallel circular disks. The calculations performed here suggest that such studies would be feasible and interesting.

\end{document}